\documentclass[twocolumn,aps,prd,nofootinbib]{revtex4-1}
\usepackage[T1]{fontenc}
\usepackage[latin1]{inputenc}
\usepackage{amssymb}
\usepackage{amsmath}
\usepackage{enumerate}
\usepackage{graphicx,wasysym}
\usepackage[pdftex]{hyperref}

\def\K{K{\"a}hler}

\makeatletter

\usepackage{epsf}

\def\be{\begin{equation}}
\def\ee{\end{equation}}
\def\ba{\begin{eqnarray}}
\def\ea{\end{eqnarray}}
\def\beas{\begin{eqnarray*}}
\def\eeas{\end{eqnarray*}}
\def\sla{\raise.15ex\hbox{$/$}\kern-.57em}

\begin{document}

\rightline{UMN--TH--3005/11, FTPI--MINN--11/15}

\title{\Large\bf Chaotic inflation and supersymmetry breaking}

\author{Renata Kallosh$^{1}$}
\author{Andrei Linde$^{1}$}
\author{Keith A. Olive$^{2}$}
\author{Tomas Rube$^{1}$}

\affiliation{$^{1}$Stanford Institute of Theoretical Physics and Department of Physics, Stanford University, Stanford, CA 94305, USA\\
$^{2}$William I. Fine Theoretical Physics Institute, University of Minnesota, Minneapolis, MN 55455, USA}


\begin{abstract}
We investigate the recently proposed class of chaotic inflation models in supergravity with an arbitrary inflaton potential $V(\phi)$. These models are extended to include matter fields in the visible sector and we employ a mechanism of SUSY breaking based on a particular phenomenological version of the KKLT mechanism (the KL model). We describe specific features of reheating in this class of models and show how one can solve the cosmological moduli and gravitino problems in this context.

\end{abstract}

\maketitle

\section{Introduction}

The simplest and most general version of the inflationary theory is the chaotic inflation scenario \cite{Linde:1983gd}. In this scenario, inflation can occur without any recourse to high temperature phase transitions, which was the trademark of old and new inflation. Chaotic inflation may occur in any model where the scalar potential is sufficiently flat, including large-field models with potentials as simple as $m^{2}\phi^{2}/2$,  $\lambda\phi^{4}/4$ and ${\lambda\over 4} (\phi^{2}-v^{2})^{2}$. However,  implementing this scenario in supergravity is a challenge.

The main difficulty in coupling chaotic inflation to supergravity is related to the \K\, potential ${ K}$. In minimal $\mathcal N=1$ supergravity, the  \K\, potential contains terms proportional to $\Phi\bar\Phi$. The F-term part of the scalar potential is proportional to $e^{ K}$, and therefore the potential scales like $e^{|\Phi|^2}$. This is much too steep for chaotic inflation at $\Phi \gg 1$.

One way to overcome this problem is to find flat directions of the inflaton potential in supergravity, see e.g. \cite{Goncharov:1985ka,Gaillard:1995az}. The simplest model of this type was proposed in Ref. \cite{Kawasaki:2000yn}. The basic idea is that instead of considering a minimal \K\, potential containing $\Phi\bar\Phi$, one may instead consider the  potential $(\Phi-\bar\Phi)^2/2$. This potential has shift symmetry: It does not depend on the field combination $\Phi+\bar\Phi$. Therefore the dangerous term $e^{ K}$ is also independent of $\Phi+\bar\Phi$, which makes the potential flat and suitable for chaotic inflation, with the field $\Phi+\bar\Phi$ playing the role of the inflaton. The flatness of the potential is broken only by the superpotential $mS\Phi$, where $S$ is an additional scalar field, which vanishes along the inflationary trajectory. As a result, the potential in the direction $\Phi+\bar\Phi$ becomes quadratic, as in the simplest version of chaotic inflation.

This work was followed by many related papers on this subject \cite{Yamaguchi:2000vm,Davis:2008fv}.  A similar idea was used in the models of chaotic inflation in string theory \cite{Silverstein:2008sg}; see \cite{Baumann:2009ni} for recent reviews.

Our present goal is to continue investigation of chaotic inflation in supergravity following the recent series of papers \cite{Kallosh:2010ug,Kallosh:2010xz}. There, it was shown that one can significantly generalize the model of Ref. \cite{Kawasaki:2000yn} by studying more general \K\ potentials of the functional form $K((\Phi-\bar\Phi)^2,S\bar S)$ and by introducing models with a superpotential ${W} = S f(\Phi)$, where $f(\Phi)$ is an arbitrary holomorphic function. In this class of models one can implement the chaotic inflation scenario with an arbitrary inflaton potential. This means that all observational results which can be successfully interpreted in the context of any phenomenological model of single-field inflation can also be obtained in the context of inflationary models based on supergravity.  Moreover, according to \cite{Demozzi:2010aj}, some models of this type lead to a natural realization of the curvaton scenario \cite{curva} with a controllable level of non-gaussianity of the adiabatic perturbations of the metric. This extends our possibilities even further.

However, a complete model of inflation in supergravity should address two additional problems. First of all, it should introduce a small amount of supersymmetry breaking at the end of inflation. It should also address the cosmological moduli problem, which plagues many cosmological models based on supergravity  \cite{Polonyi}. This is a generic problem for models based on the simplest mechanism for breaking supersymmetry using a linear superpotential \cite{pol}. One may try to solve this problem with a generalization of the Polonyi potential in some of the new cosmological models \cite{Takahashi:2010uw} using the adiabatic relaxation mechanism proposed in \cite{Linde:1996cx}. Indeed, our investigation of this issue suggests that this mechanism does work for certain versions of our scenario and we discuss this briefly in section III . Alternatively, one may turn to non-minimal models based on no-scale supergravity \cite{cremmer}. While these models can successfully stabilize one of the two flat directions associated with supersymmetry breaking \cite{ekn,msy2,abdk2}, one of the flat directions is left unfixed.

While no-scale supergravity may be a step in the right direction, if one wants to consider string theory inspired versions of supergravity, one may need to take into account some unusual but rather generic features of string cosmology based on the KKLT mechanism of vacuum stabilization \cite{Kachru:2003aw}.  First of all, supersymmetry breaking is a generic feature of the string theory models with vacuum stabilization, which may make other mechanisms of supersymmetry breaking redundant. Secondly, supersymmetry breaking in this class of models cannot be attributed to the F-term alone; one should take into account the effect of the uplifting of the potential required for the fine-tuning of the cosmological constant. One may interpret uplifting either as a soft supersymmetry breaking induced by string theory effects \cite{Kachru:2003aw}, or as a D-term contribution \cite{Burgess:2003ic,Achucarro:2006zf}. In this class of models, the Hubble constant during inflation typically must be smaller than the gravitino mass, unless one does something special, e.g. fine-tunes the superpotential of the volume modulus in a specific way (the KL mechanism) \cite{Kallosh:2004yh}.  As we will see, once both of these effects are taken care of, the cosmological moduli problem disappears even without the use of the mechanism proposed in \cite{Linde:1996cx,Takahashi:2010uw}.

To complete our construction of the inflationary scenario based on supergravity  we need to construct the theory of reheating in this scenario. As we will see (see also \cite{Endo:2007sz,Takahashi:2010uw}), reheating in the theories with flat directions has some distinguishing features which we are going to analyze. In particular, the reheating temperature in this class of models is naturally suppressed~\cite{ekoty}, which simplifies the solution of the cosmological gravitino problem \cite{gravitino,nos}. 

In what follows, we will first describe our approach to general inflationary potentials based on $\mathcal N=1$ supergravity, see Section II. In section III, we discuss the mechanism of supersymmetry breaking.
After briefly reviewing past (more traditional) approaches, we describe the KL model
along with some of its phenomenological consequences.  In section IV, we introduce 
a combined theory of inflation and supersymmetry breaking and describe the evolutionary behavior
of our 3-field system.  Reheating in this class of models is discussed in section V, and our conclusions
are summarized in section VI.

\section{General Inflationary Potential}\label{generalpot}
This section reviews the supergravity theory of inflation of \cite{Kawasaki:2000yn,Kallosh:2010ug,Kallosh:2010xz}. The inflaton sector consists of two fields: the inflaton field, $\Phi$, and the stabilizer field, $S$. The real part of the field $\Phi$ will play the role of the inflaton. Meanwhile, the fields $S$ and ${\rm Im}\, \Phi$ will be forced to vanish during inflation. The scalar potential for uncharged chiral superfields in $\mathcal N=1$ supergravity is
\be
V = e^G \left( G_i G^{i \bar j}G_{\bar j} -3 \right), 
\label{scalarpot}
\ee
or using 
\be
G = { K} + \log |{ W}|^2 \ ,
\ee
\be
V=e^{ K}\left({ K}^{i\bar j}D_i W\bar D_{\bar j}\bar{ W}-3| W|^2\right),
\label{eqn:SUGRApotential}
\ee
where $D_i W\equiv\partial_i W+{ K}_iW$. For generic \K\, potentials, the exponential renders the potential far too steep for inflation. One way of getting around this problem is to impose a shift symmetry on $\Phi$. The \K \, potential is for simplicity chosen to have functional form
\be
 K((\Phi-\bar\Phi)^2,S\bar S)
\ee
The shift symmetry not only flattens the potential along the real $\Phi$ direction, but by rescaling the fields, the field metric can be chosen to be canonically normalized along the inflaton path $S={\rm Im}\,\Phi=0$: ${ K}_{S\bar S}={ K}_{\Phi\bar\Phi}=1$. Furthermore, using a \K\, transformation, $ K$ can be made to vanish along this path.

The superpotential is chosen to be
\be
{W}= Sf(\Phi) \ ,
\label{cond}
\ee
where $f(\Phi)$ is a real holomorphic function such that $\bar f(\bar \Phi) = f(\Phi)$. Any function which can be represented by Taylor series with real coefficients has this property.  This superpotential has a number of good properties. First, both  $ W$ and $D_\Phi W$ vanish at $S=0$. As such, the only non-vanishing contribution to the scalar potential comes from $F_S=D_S W=f(\Phi)$. Along the inflaton's trajectory where $\text{Im}\Phi=0$, we obtain the amazingly simple potential
\be
V=|f(\Phi)|^2.
\ee
Second, the superpotential and  \K\, potential are odd and even respectively,  under the transformation $S\rightarrow -S$. Looking at (\ref{eqn:SUGRApotential}), we see that this makes the scalar potential invariant and that $S=0$ is, therefore, an extremum. Finally, the reality condition implies that both $|f(\Phi)|^2$ and $K((\Phi-\bar\Phi)^2,S\bar S)$ are invariant under $
\Phi\rightarrow\bar\Phi$, making ${\rm Im}\,\Phi=0$ an extremum.

Next consider the stability of potential with respect to transverse perturbations. Using the basis 
\be
S=\frac{1}{\sqrt{2}}(s+i\alpha),\quad \Phi=\frac{1}{\sqrt{2}}(\phi+i\beta).
\ee
the inflaton potential becomes
\be
V(\phi)=f^2(\phi/\sqrt{2}).
\ee
The masses of these fields were calculated in \cite{Kallosh:2010xz} and found to be
\ba\label{genstab2}
m^2_{\beta} &=&V \left[ 2  (1
-  K_{\Phi\bar\Phi S\bar S})+2\epsilon -\eta\right]\ , \\
m^{2}_{s} &= & m^{2}_{\alpha}= V\left[-K_{S\bar S S\bar S}+\epsilon\right] \ .
 \label{masss}\ea
where
\be
\epsilon=\frac{1}{2}\left(\frac{\partial_\phi V}{V}\right)^2=\frac{(\partial_\Phi f)^2}{f^2},\,\,
\eta=\frac{\partial_\phi^2V}{V}=\frac{\partial_\Phi^2 f}{ f}+\frac{(\partial_\Phi f)^2}{f^2}
\ee
are slow roll parameters. The degeneracy is explained by the unbroken R-symmetry $S\rightarrow e^{2i\alpha }S$. The dependence on ${K}_{\Phi\bar\Phi S\bar S}$ and ${ K}_{S\bar S S\bar S}$ can be understood by noting that the inflaton potential is generated by the F-term of the $S$ field and the corresponding field metric is
\be
{ K}^{S\bar S}=\frac{1}{1+2{ K}_{\Phi\bar\Phi S\bar S}\beta^2+{ K}_{S\bar S S\bar S}(s^2+\alpha^2)+...}.
\ee
The condition for a transverse scalar field to remain fixed despite quantum fluctuations during inflation is $m_\perp^2\gtrsim H^2$. If the potential supports slow roll inflation both $\epsilon$ and $\eta$ are tiny and can be dropped. Using $V=3H$ we get a condition on $ K$:
\be
{ K}_{\Phi\bar\Phi S\bar S}\lesssim\frac{5}{6},\quad
{ K}_{S\bar S S\bar S}\lesssim-\frac{1}{3}.
\ee
Note that the stability condition is independent of the details of the inflaton potential as long as the slow roll parameters are sufficiently small. After inflation, $\epsilon$ and $\eta$ grow and at the minimum of the potential, where $V=V'=0$, the masses are
\be
m_\phi^2=m_\alpha^2=m_s^2=m_\beta^2=\partial_\phi^2V.
\label{eqmass}
\ee

As long as the inflationary trajectory is stabilized, the explicit expression for the \K\, potential does not play any role for inflation. However, it is helpful to consider some particular examples.

One may consider a simple polynomial \K\, potential \cite{Kallosh:2010ug,Kallosh:2010xz}  that is a generalization of the potential used in \cite{Kawasaki:2000yn,Yamaguchi:2000vm,Davis:2008fv}:
\begin{equation}
{K} =  S \bar S - \frac{1}{2}(\Phi -\bar\Phi)^2 - {\zeta} (S\bar S)^2  +  \frac{\gamma}{2} S\bar S (\Phi -\bar\Phi)^2.
\label{K2}
\end{equation}
Note that the stabilizing terms $- {\zeta} (S\bar S)^2  + {\gamma\over 2} S\bar S (\Phi -\bar\Phi)^2$ were added to the \K\, potential of the model of \cite{Kawasaki:2000yn,Yamaguchi:2000vm,Davis:2008fv}. This \K\ geometry has $K_{\Phi\bar\Phi S\bar S}= -\gamma$ and $K_{S\bar S S\bar S}=-4\zeta$ and the stability conditions during inflation are, for any sufficiently flat $f(\Phi)$, $\gamma \gtrsim -5/6$ and $\zeta\gtrsim 1/12$. 

Another example is the logarithmic \K\, potential \cite{Kallosh:2010ug,Kallosh:2010xz}, that is a generalization of the potential used in \cite{Einhorn:2009bh,Ferrara:2010yw,Lee:2010hj,Ferrara:2010in}:
\ba
{K} &=& -3\log \Bigl[ 1 + \frac{1}{6}(\Phi -\bar\Phi)^2 - \frac{1}{3}S \bar S +\zeta (S\bar S)^2/3  \nonumber\\ &-&   \frac{\gamma}{6} S\bar S (\Phi -\bar\Phi)^2\Bigr] .
\label{Ka}
\ea
In this case $K_{\Phi\bar\Phi S\bar S} = -\gamma+1/3$ and $K_{S\bar S S\bar S}=-4\zeta+2/3$ and the stability conditions with respect to the generation of inflationary perturbations of the fields orthogonal to the inflationary trajectory are $\gamma \gtrsim -1/2$ and $\zeta \gtrsim 1/4$.

\section{The scale of inflation and SUSY breaking: KL model}

In the inflationary model discussed so far, supersymmetry is unbroken in the vacuum state corresponding to the minimum of the potential with $V = 0$. There are several ways to introduce supersymmetry breaking to this model. 
The simplest way is to add the Polonyi field $z$ with a linear superpotential $W = \mu (z + b)$
and ${K} = z \bar{z}$ \cite{pol}. For the choice
$b = 2- \sqrt{3}$, the scalar potential has a supersymmetry breaking Minkowski minimum. The gravitino mass is $m_{3/2} = e^{G/2} = e^{2-\sqrt{3}} \mu$.
Therefore, to solve the hierarchy problem, one must tune $\mu$ to $\mathcal{O}(10^{-15})$. As is well known, this theory is plagued with a moduli problem \cite{Polonyi} as
the two scalars remain light and will eventually dominate the energy density of the universe after inflation. Economically, it would be nice to be able to associate the stabilizer field $S$ with
the Polonyi field $z$, but we were unable to find a successful model of this type.

It is in principle possible to relieve this problem by modifying the theory so that 
the field $z$ obtains a large mass during inflation and adiabatically relaxes to its minimum \cite{Linde:1996cx}.  For example,  one could add a quartic term to the \K\ potential ${K} = z \bar{z} + R\, (z \bar{z})^2$ and also add a quadratic term  to the superpotential
so that $W = \mu (b + z + c z^2)$.  For given values of $R$ and $c$, $b$ must
be fine-tuned to recover a Minkowski vacuum.  For  large $R$, adiabatic
relaxation will occur  \cite{Linde:1996cx,Takahashi:2010uw}.

Another alternative is to begin with a \K\ potential of the no-scale form.
For example, the \K\ potential 
${K} = -3 \log (c + z + \bar{z} + b(z + \bar{z})^4 -  \Phi \bar{\Phi}/3)$
will fix the real part of $z$, and generate a large gravitino mass (for large $c$)~\cite{ekn}.
However this formalism is only suited for small field inflation, see e.g. \cite{eenos}.
Another choice  is ${K}  =  -3 \log (z + \bar{z}  - \Phi \bar{\Phi})
 + (1+ \kappa_S S \bar{S} + \kappa_z
(z + \bar{z}  - \Phi \bar{\Phi})) S \bar{S}$ \cite{abdk2}.
This model allows for generalized inflationary potentials of the type discussed in the previous
section, and fixes the combination $z + \bar{z}  - \Phi \bar{\Phi}$.
However,  these theories leave behind a (near) massless degree of freedom
associated with Im $z$. 

String theory suggests another approach to supersymmetry breaking, which we are going to pursue in this paper. 
In string theory, one must consider stabilization of the volume modulus $\rho$ to explain why our universe is 4d rather than 10d. The simplest approach to this issue is based on the KKLT mechanism \cite{Kachru:2003aw}. In this theory, one first finds a stable supersymmetric vacuum with a negative vacuum energy density $V_{\rm AdS}$, and then uplifts it until its vacuum energy becomes positive but negligibly small, about $10^{{-120}}$ in Planck units. After the uplifting, supersymmetry breaks down, and the gravitino mass has a simple relation to the depth of the original AdS minimum  \cite{Kallosh:2004yh}:
\be
m^{2}_{3/2} = |V_{\rm AdS}|/3 \ .
\ee
Thus the mechanism of supersymmetry breaking is built into the new generation of string theory models. One can add to it other mechanisms of supersymmetry breaking, such the Polonyi mechanism \cite{pol}, dynamical supersymmetry breaking \cite{Intriligator:2006dd}, an O'Raifeartaigh mechanism \cite{O'Raifeartaigh:1975pr,Kallosh:2006dv} or something else. However, this would make the models more complicated. Therefore, in this paper we will concentrate on the string theory based mechanism of supersymmetry breaking.

One should note that in the simplest versions of the KKLT construction a rather unusual problem has to be addressed:
 the Hubble constant during inflation cannot be greater than the gravitino mass, $H \lesssim m_{3/2}$  \cite{Kallosh:2004yh}. The reason is that in the simplest KKLT models, the barrier separating the stabilized dS vacuum from the 10d Minkowski vacuum has a height proportional to $m^{2}_{3/2}$. When the inflationary potential is added to the system, it may lift the dS minimum above the barrier. If this happens, the universe decompactifies and becomes 10-dimensional.

\begin{figure}[ht!]
\centering
\includegraphics[scale=0.72]{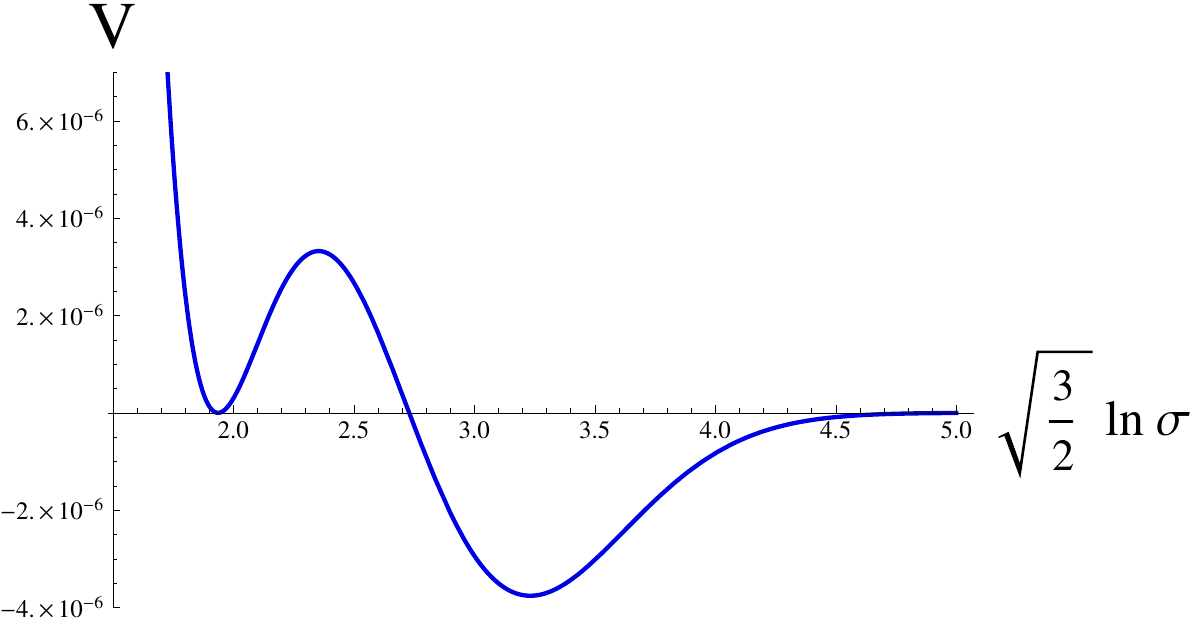}
\caption{Scalar potential of the KL model for the values of the parameters $A=B =1$, $a = \pi/25$, $b = \pi/10$ as a function of the canonically normalized volume modulus  field $\sqrt{3/2} \ln \sigma$.}
\label{KLpotential}
\end{figure}

One can try to solve this problem in several different ways, see for example \cite{Kallosh:2004yh,Davis:2008fv,Badziak:2008yg,Conlon:2008cj}.
The simplest mechanism involves a slightly generalized KKLT model, which is sometimes called the KL model \cite{Kallosh:2004yh}. In this model,  the \K\, potential of the volume modulus $\rho$ describing the size of compactification is
\be
{K}_\text{KL} = - 3 \ln[(\rho + \bar{\rho})]
\ee
and instead of the standard KKLT superpotential $W = W_0 + Ae^{-a\rho}$, one uses the racetrack superpotential 
\be
W_\text{KL} = W_0 + Ae^{-a\rho}- Be^{-b\rho} \ .
\label{adssup}
\ee
For a particular choice
\be\label{w0}
W_0= -A \left({a\,A\over
b\,B}\right)^{a\over b-a} +B \left ({a\,A\over b\,B}\right) ^{b\over b-a} ,
\ee
the potential $V(\rho)$ has a  supersymmetric Minkowski  minimum at ${\rm Im } \rho = 0$ and
\be
 \sigma_{0}= {1\over a-b}\ln \left ({a\,A\over b\,B}\right)\, ,
\label{sigmacr} \ee
where $\sigma$ is a real part of the field $\rho$.
In this minimum 
\be
W(\sigma_{0})=0 \ , \qquad D_\rho W(\sigma_{0})=0 \ , \qquad V(\sigma_{0})=0 \ .
\label{susy} \ee
The shape of the potential, $V$, for a particular set of parameters $A=B = 1$, $a = \pi/25$, $b = \pi/10$,  is shown in Fig. \ref{KLpotential},  as a function of the canonically normalized volume modulus  field $\sqrt{3/2} \ln \sigma$.

One can show that because of the relations $W(\sigma_{0})=0$, $D_\rho W(\sigma_{0})=0$, the mass squared of the field $\sigma$ at the minimum of the potential with $V = 0$, as well as the mass squared of the imaginary component of the field $\rho$,  is given by ${2\over 9}\sigma_{0}\, W^{2}_{\rho,\rho}(\sigma_{0})$. In the KL model, one finds 
\be
m^{2}_{\sigma} = \frac{2}{9} a\, A\, b\, B\, (a-b) \left(\frac{a A}{b B}\right)^{-\frac{a+b}{a-b}}\,  \ln\left (\frac{a A}{b B}\right) \ .
\ee 
For the particular choice of parameters 
\be
A=B =1,\,a = \pi/25,\,b = \pi/10
\label{eqn:paramset}
\ee
one has $m_{\sigma} \sim 1.3\times  10^{-2}$, in Planck units, so it is typically much heavier than the inflaton field, which, in the simplest model of chaotic inflation has mass $m_{\phi} \sim 6 \times 10^{{-6}}$. Thus the problems discussed above associated with fixing {\em both} components of the Polonyi-like field are resolved in the KL model. This hierarchy of mass scales is one of the necessary conditions which is required to ignore the dynamics of the volume modulus $\sigma$ during inflation. More exact requirements will be discussed below.

It will be useful to understand the properties of the KL potential and the mass of the volume modulus under the simultaneous  rescaling of the parameters $A \to C A$, $B\to CB$. This rescaling does not affect the position of the minimum $\sigma_{0}$, but  it increases the value of $W_{0}$ and the mass of the volume modulus  by a factor $C$, and it increases  the height of the barrier in the KL potential by a factor $C^2$. 
Meanwhile the simultaneous rescaling $a \to c a$ and $b \to c b$  decreases $\sigma_{0}$ by a factor of $c$, increases $m_{\sigma}$ by a factor of  $c^{{3/2}}$, and increases the height of the barrier by a factor of $c$. These facts will be important for our discussion of moduli stabilization during inflation in the context of this scenario.

In the KL model discussed so far, supersymmetry is unbroken in the vacuum state corresponding to the minimum of the potential with $V = 0$. The scale of  supersymmetry breaking  will be determined by a slight perturbation of the superpotential (\ref{adssup}) by adding to it a small constant $\Delta W \propto \mu$. Independent of the sign of $\Delta W$, the constant shifts the minimum of the potential $V$ from zero to its negative value $V_{\rm AdS} < 0$. Therefore $V_{\rm AdS}$ in the first approximation must be proportional to $-\Delta W^{2}$. After some algebra, one finds that the position of the minimum shifts from $\sigma_{0} $ by $\Delta\sigma = {3\Delta W\over 2\sigma_{0} W_{\rho,\rho}}$, and the potential at the minimum becomes
\be
V_{\rm AdS}(\Delta W) = -{3(\Delta W)^2\over 8 \sigma_{0}^{3}}= - {3\over 8} \left({a - b \over  \ln \left({a A\over b B}\right)}\right)^{3}\, (\Delta W)^2 \ .
\label{vads}
\ee 
In this minimum, the value of the superpotential (including the additional constant $\Delta W$), remains equal to $\Delta W$ up to small corrections $O(\Delta W)^2$. Supersymmetry in the minimum is still unbroken, $D_\rho W = 0$, whereas $W_{\rho} = {3\over 2 \sigma_0}\Delta W$. 

Uplifting of the AdS minimum induces supersymmetry breaking and is achieved by adding to the potential a term 
\be
\Delta V \approx |V_{\rm AdS}(\Delta W)|\ {\sigma^{n}_{0}\over\sigma^{n}}.
\ee
 In the original KKLT construction it was assumed that $n = 3$ \cite{Kachru:2003aw}, but according to \cite{Kachru:2003sx} $n = 2$ in the uplifting term, due to effects related to warping. One may have $n = 3$ if the uplifting occurs due to a D-term \cite{Burgess:2003ic,Achucarro:2006zf}.  Because of the dependence of the uplifting term on $\sigma$, the minimum after the uplifting shifts to slightly greater values of $\sigma$. However, this effect is extremely small, being proportional to $(\Delta W)^{2}$. Therefore, as a first approximation, the position of the minimum, as well as the values of $W$ and of its first derivative $W_{\rho}$, remain the same as they were before uplifting, independent of $n$.

After uplifting to the present state with a nearly vanishing vacuum energy, the gravitino mass  becomes 
\be
m_{3/2} = \sqrt{|V_{\rm AdS}|/3} =  {1\over 2\sqrt 2}\left({a - b \over  \ln \left({a A\over b B}\right)}\right)^{3/2}\, |\Delta W| \ .
\ee 
In particular, for $A=B =1$, $a = \pi/25$, $b = \pi/10$, one has $m_{3/2} \sim 3 \times 10^{-2} |\Delta W| \sim 2.3\, m_{{\sigma}} |\Delta W|$.  To have  $m_{{3/2}} \sim 1$~TeV, which is about $0.4\times 10^{{-15}}$ in Planck units, one should have $|\Delta W| \sim 10^{-14}$. This means that to make the gravitino mass comparable to the electroweak scale, we must introduce a small parameter $\sim 10^{-14}$, which is comparable to the small parameter, $\mu$, required in the standard Polonyi superpotential. This should be done in the context of a special version of the KKLT theory which, in the state with $\Delta W = 0$, describes a supersymmetric Minkowski vacuum. Note that in this class of models, unlike in the simplest KKLT models, the mass of the volume modulus, as well as the inflaton mass, can be many orders of magnitude greater than the gravitino mass, and the light Polonyi field is not required for supersymmetry breaking. This solves the cosmological moduli problem in this class of models.

It would be interesting to find out how generic models of this type are in the landscape. We do not have a complete answer to this question; certainly these models are fine-tuned. However, we would like to mention  an interesting aspect of this class of models revealed in \cite{BlancoPillado:2005fn}.

Vacuum stabilization in string theory is quite complicated because one should achieve stability with respect to {\bf all} string theory moduli. This problem was solved for a particular class of models, see e.g. \cite{Denef:2005mm}, but it is certainly true that the requirement of stability with respect to all  moduli is a significant constraint, limiting the total number of stable string theory vacua. In this respect,  it is interesting that all Minkowski vacua  with unbroken supersymmetry are stable automatically  \cite{BlancoPillado:2005fn}, due to the positive energy theorem in supergravity \cite{Deser:1977hu}.

We would like to go beyond this simple statement and find what happens when one introduces supersymmetry breaking in the KL model.  Let us first analyze the second derivative of the  F-term potential $V$, Eq. (\ref{eqn:SUGRApotential}), for all scalars in supergravity at the supersymmetric minimum in terms of a covariantly holomorphic complex gravitino mass 
$
e ^{ { K}/2}W \equiv {\it m}(z, \bar z) 
$,  related to the (real) gravitino mass,
$
 m^2_{3/2}  =|m \, \bar m|= e^G
$. Note that the complex masses of the chiral fermions in $\mathcal N=1$ supergravity are equal to
\begin{equation}
  D_i D_j {\it m} \equiv {\it m}_{ij}\ , \qquad \bar D_{\bar i}\bar D_{\bar j}\bar {\it m}\equiv \bar {\it m}_{\bar i  \bar j} \ .
\label{chiral}
\end{equation}
At the supersymmetric minimum  $\partial _i V=0$ and 
\be
D_i {\it m}  \equiv {\it m}_i =0 \ , \qquad \bar D_{\bar i} \bar {\it m}  \equiv \bar {\it m}_{\bar i}=0\ .   
\ee
In a supersymmetric Minkowski minimum, where $m=\bar m=0$, the matrix of the second derivatives of the potential $V$  is positive definite,
\begin{equation}
  \partial_{i} \partial _{\bar j} V|_{Mink}=  {\it m}_{ik}{ K}^{k\bar k} \,\bar {\it m}_{\bar k \bar j}= |m_{i{\bar j}}|^2 \geq 0 \ ,
\label{mink}
\end{equation}
in agreement with the general stability expectations based on \cite{Deser:1977hu}. This is exactly what we found in the particular version of the KL model studied above, with the masses of the real and imaginary part of the field $\rho$ being quite large, $O(10^{{-2}})$ in Planck units.

If we modify this model and add the term $\Delta W$, the Minkowski minimum becomes a supersymmetric AdS extremum, with  the second derivatives of the potential at $\partial _i V=0$ given by
\ba  \partial_{j} \partial _i V &=&-{\it m}_{ji} \bar {\it m}\ , \qquad \partial_{\bar j} \partial _{\bar i} V =-\bar {\it m}_{\bar j \bar i} {\it m} \ , \nonumber \\
\partial_{\bar j} \partial _i V&=& -2 K_{\bar j i} {\it m} \bar {\it m} + {\it m}_{ik}K^{k\bar k} \,\bar {\it m}_{\bar k \bar j} \ .
\label{mix}
\ea
This mass matrix differs from Eq. (\ref{mink}) by small terms proportional to the gravitino mass. 

In the simplest versions of the KKLT model,  the mass of the volume modulus typically is of the same order as $m_{3/2}$. That is why vacuum stability in these models is not automatic, and the situation may become even more complicated after the uplifting, see for example \cite{Choi:2004sx} and references therein.

In this respect, the situation in the KL model is much better. If the mass matrix in the Minkowski vacuum is positive definite, $|m_{ij}|^2 > 0$, i.e. if it is a minimum, then it should remain a minimum of the scalar potential after adding the term $\Delta W$, if the gravitino mass $m_{3/2} \sim \Delta W$ is much smaller than $|m_{ij}|$. In the particular model considered above this condition is easily satisfied. 
This result is unchanged by uplifting. The uplifting term depends only on the field $\sigma$, which is strongly stabilized near $\sigma_{0}$. As we already mentioned, after uplifting, the field $\sigma$ in the KL model remains practically unchanged; its modification is suppressed by $(\Delta W)^{2}$, which in our case is $O(10^{{-28}})$. That is why the second derivatives of the potential $V$  after uplifting remain the same as before uplifting: The potential is simply shifted upwards without changing its shape {\it with respect to all moduli fields}. This means that if the gravitino mass is sufficiently small,  supersymmetry breaking in the KL model does not destabilize the potential  \cite{BlancoPillado:2005fn}. 

This suggests that at least some part of the fine-tuning involved in the formulation of the KL-type models is justified by vacuum stability which is much easier to achieve in these models. Moreover, these considerations hint towards a possible reason for the smallness of supersymmetry breaking: The smaller is the gravitino mass, the easier it is to stabilize the vacuum in this class of string theory models.

A more detailed study of vacuum statistics in string theory landscape is required to evaluate potential significance of these arguments. But quite independently of these considerations, we already found that this class of models has an important advantage that we would like to reemphasize:  
They help solve the long-standing cosmological problem associated with light moduli fields, such as the Polonyi field. Such fields typically accumulate  a significant amount energy and decay too slowly, which leads to disastrous cosmological consequences \cite{Polonyi}. In the KL model, this problem does not appear because we do not need to have light Polonyi fields;  supersymmetry breaking is associated with the volume modulus. In the KL models, this field is superheavy by construction.

Before returning to our central question of inflation, 
we comment on the phenomenology induced by the KL model.
If we extend the theory to include a minimally coupled matter sector, our \K\, potential becomes
\be
K = - 3 \ln[(\rho + \bar{\rho})] + y^i \bar{y_i}
\ee
along with the superpotential
\be
W = W(\rho) + W_{SM}(y_i) \ ,
\ee
where we include standard model fields, $y_i$ and $W_{SM}(y_i)$ is the 
Standard Model superpotential (we are using $y_i$ to denote both the scalar component
and superfield). The scalar potential is given by
\be
V_{SM}  =  \frac{e^{|y_i|^2}}{8\sigma_0^3} \left( \left| \frac{\partial W_{SM}}{\partial y^i} + \bar{y_i} 
W \right|^2 + 3 \left| W_{SM} \right|^2 - 3 \left| W \right|^2 \right),
\ee
where $W = W_{SM} + \Delta W$ and we assume a sum over Standard Model fields). In the low energy limit, the potential at the uplifted minimum (i.e. subtracting the contribution from Eq. \ref{vads}), can be written as
\begin{eqnarray}
V_{SM} & = & \frac{1}{8\sigma_0^3} \left( \left|\frac{\partial W_{SM}}{\partial y^i}\right|^2 
+ (\Delta W)^2 y^i \bar{y_i}  \right. \\ \nonumber
& & \left. + \left[ (\Delta W) (\bar{y_i} \frac{\partial W_{SM}}{\partial y^i} - 3 W_{SM}) + h.c. \right]  \right),
\end{eqnarray}
or 
\begin{eqnarray}
V_{SM} & = &   \left|\frac{\partial W_{SM}}{\partial y^i}\right|^2  +m_{3/2}^2 y^i \bar{y_i} \\ \nonumber
& & +\left[ m_{3/2} ( \bar{y_i} \frac{\partial W_{SM}}{\partial y^i} - 3 W_{SM})+ h.c.\right ] 
\end{eqnarray}
after a rescaling of the superpotential $W_{SM} \to 2\sqrt{2} \sigma_0^{3/2} W_{SM}$.
This corresponds to a standard minimal supergravity model (mSUGRA) with a
universal scalar mass, $m_0 = m_{3/2}$ and
a trilinear supersymmetry breaking $A$-term, $A_0 = 0$ (and a bilinear term $B_0 = -m_0$).
This is distinct from the prediction of the Polonyi model where $A_0 = (3 - \sqrt{3}) m_0$ and $B_0 = 
(2 - \sqrt{3}) m_0$ or no-scale supergravity with $A_0 = B_0 = m_0 = 0$.

\section{Inflation in combined theory}
Next we study the inflation potential in the combined theory
\begin{align}
 K&={ K}_\text{inf}((\Phi-\bar\Phi)^2,S\bar S)-3\text{log}(\rho+\bar \rho) \ ,\\
 W&=Sf(\Phi)+{W}_0+Ae^{-a\rho}-B e^{-b\rho} +\Delta W \ .
\end{align}
Supergravity couples the inflaton and KL sectors to each other. In this section, we discuss how the inflation model in section \ref{generalpot} is affected by the KL sector. In the simplest string theory inflation models based on the KKLT scenario, the energy stored in the inflaton potential can destabilize the volume modulus if it is too large, leading to the constraint $H \lesssim m_{{3/2}}$ \cite{Kallosh:2004yh}. In the KL model, this constraint disappears because the height and steepness of the stabilization potential are not related to the gravitino mass, so they can be very large. A useful quantity that parametrizes the relative size of the inflaton potential and the KL-barrier is
\be
\delta=-\frac{3\sqrt{3}}{4}\frac{f(\phi/\sqrt{2})}{\sigma_0^2{ W}_\text{KL}''} .
\ee
For the parameters in (\ref{eqn:paramset}) this is $\delta\sim 4.3f(\phi/\sqrt{2})\ll1$.  One may also consider the models with large SUSY breaking and large gravitino mass, as in \cite{Davis:2008fv}, but we will follow the conventional route and assume that the gravitino mass is many orders of magnitude smaller than the Hubble constant during inflation. For this reason, we can neglect the term $\Delta W$ in our investigation of inflation.

In addition to stabilization of the volume modulus, one should take care of the stability of the fields $S$ and ${\rm Im} \, \Phi$. If the masses of some of these fields at $S={\rm Im} \, \Phi =0$ are smaller than $H$, then they can easily shift away from the origin, and the resulting inflationary evolution becomes very complicated  \cite{Davis:2008fv}.  In some cases, this may lead to undesirable isocurvature perturbations, or to the realization of the curvaton scenario \cite{curva}, as recently discussed in \cite{Demozzi:2010aj}. In our investigation, we will try to find a regime such that the volume modulus $\rho$ is strongly stabilized near the minimum of the KL potential at $\rho = \sigma_{0}$ and the fields $S$ and ${\rm Im} \, \Phi$ are strongly stabilized near their zero values. In this case, the inflaton field will travel along the real $\Phi$ axis with $S\approx 0$ and $\rho\approx \sigma_0$, and inflation will occur just like in the single field chaotic inflation model with the potential $V = f^{2}(\phi/\sqrt 2)$ \cite{Kallosh:2010ug,Kallosh:2010xz}.
\begin{figure}[t]
\centering
\includegraphics[width=8.5cm]{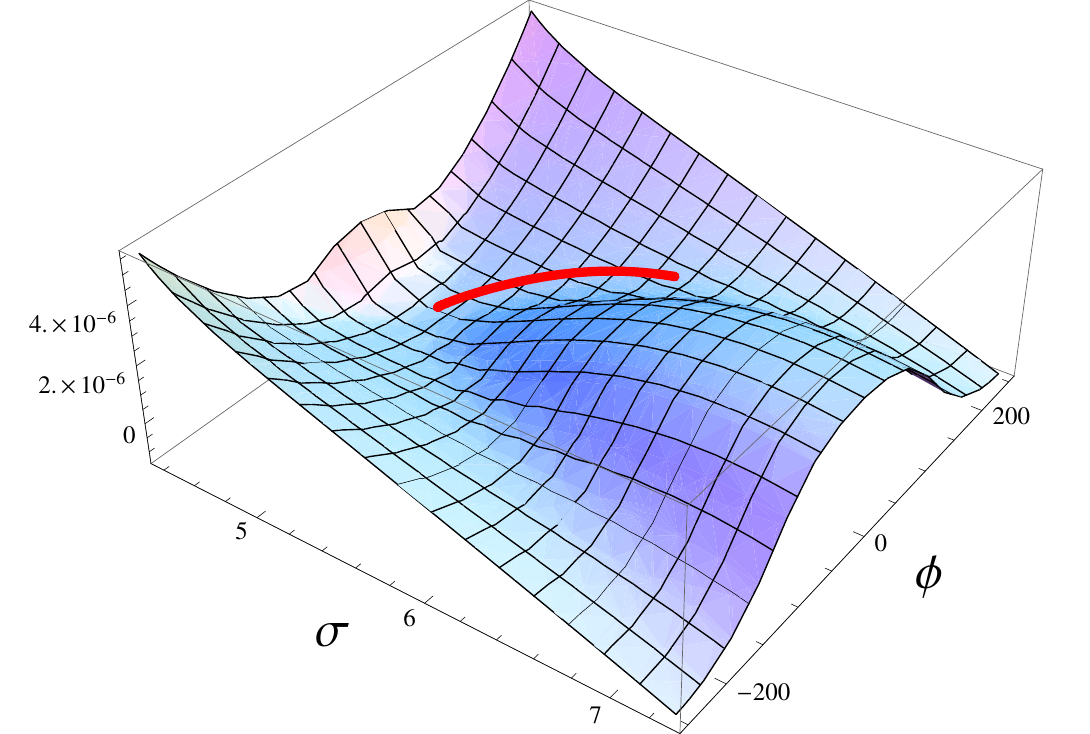}
\includegraphics[width=8.5cm]{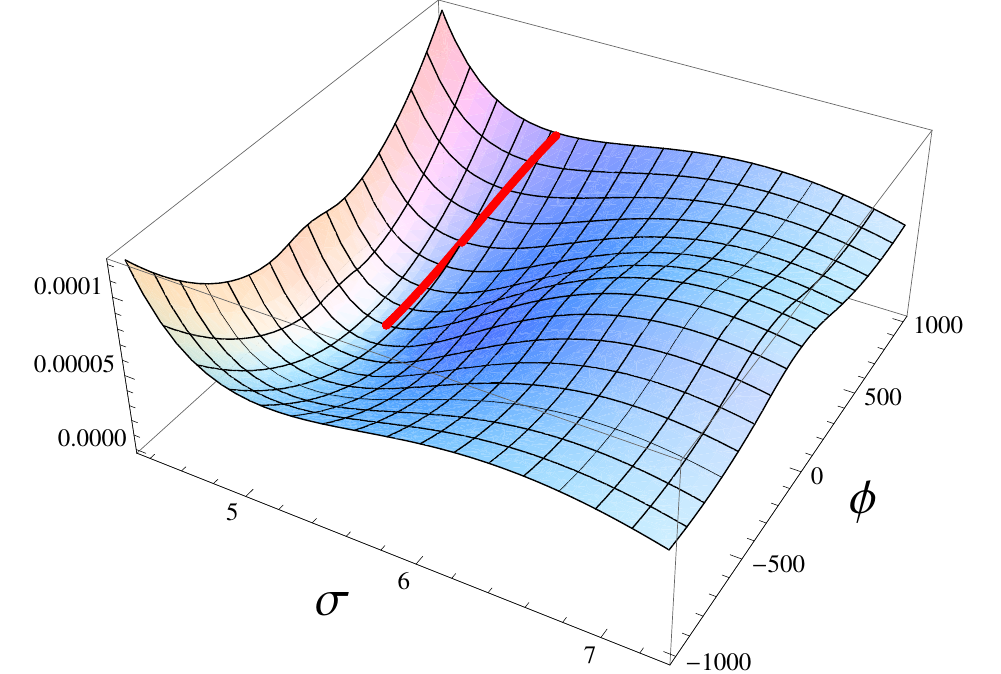}
\caption{The first of these plots shows the potential with the parameters in (\ref{eqn:paramset}) as a function of $\phi$ and $\sigma$. The \K \,potential has $ K_{S\bar S S\bar S}=0$ (no stabilization of $s$, as in \cite{Davis:2008fv}). For each value of $\phi$ and $\sigma$, $s$ is adjusted so that the potential is minimized.  The red line shows the inflaton trajectory.  Inflation is possible for $|\phi|< 140$; further increase of $\phi$ destabilizes the potential. 
The second plot corresponds to the situation where we stabilize $s$ near $s = 0$ by taking $ K_{S\bar S S\bar S}=-4$. We increase $A$ and $B$ by a factor of 5, to increase the height of the barrier. In this regime inflation is possible for $\phi$ well above 1000, in Planck units. In the investigation of the observational consequences of inflation in this regime one can ignore the KL potential and use the results of the previous investigation of chaotic inflation in supergravity \cite{Kallosh:2010ug,Kallosh:2010xz}. }
\label{fig:CombinedPotential}
\end{figure}

Supergravity introduces a number of couplings between the inflation and KL sector. The scalar potential can be decomposed as
\be
V=e^{{ K}_\text{inf}}V_{KL}+\frac{V_\text{inf}+3e^{K_\text{inf}}|W_\text{inf}|^2}{(\rho+\bar\rho)^3}+V_\text{mix}.
\label{eqn:mixedpotential}
\ee
The subscripts "inf" and "KL" denote the potentials in the decoupled limits studied in previous sections and
\begin{align}
V_\text{mix}=&\frac{e^{{ K}_\text{inf}}}{(\rho+\bar\rho)^3}\left(
{ K}^{a\bar b}{ K}_a K_{\bar b}|W_\text{KL}|^2
\right.\cr
&\left.+2\,\text{Re}\left[{ K}^{a\bar b}{ D}_a{ W}_\text{inf}{ K}_{\bar b}\bar { W}_\text{KL} \right. \right. \cr
&\left. \left. - (\rho+\bar\rho){ W}_{\text{KL},T}\bar{ W}_\text{inf} \right] 
\right) . 
\end{align}
This potential has a number of interesting features. First, $V_\text{mix}$ vanishes at $\rho=\sigma_0$ since ${ W}_\text{KL}={ W}_{\text{KL},T}=0$. Secondly, the $3|W_\text{inf}|^2$-term in (\ref{eqn:mixedpotential}) is due to the no-scale form of ${K}_\text{KL}$:
\be
{ K}^{T\bar T}|{K}_T W|^2\supset { K}^{T\bar T}{ K}_T{ K}_{\bar T}|{ W}_\text{inf}|^2=3|{ W}_\text{inf}|^2.
\label{eqn:NoScale}
\ee
This cancels the $-3| W|^2$ term in (\ref{eqn:SUGRApotential}) and the inflaton sector thus inherits the no-scale structure of the KL sector. Finally, during inflation, the inflaton won't exactly follow $S=0$, $\sigma=\sigma_0$ but be displaced transversely by a small amount $\{\delta s,\delta \sigma\}$. The biggest effect comes from the term $V_\text{inf}/(\rho+\bar\rho)^{3}$ which, if it becomes large enough, can destabilize the KL-barrier.  As long as the perturbation is small the new minimum can be found by expanding the potential to quadratic order and solving for the displacement that puts the first derivative to zero. The first derivative along the unperturbed trajectory is
\be
\left(\begin{array}{c}\partial_\phi\cr\partial_s \cr\partial_{\tilde\sigma}\end{array}\right) V(\phi,0,\sigma_0)=\left(\begin{array}{c}\sqrt{2\epsilon}\cr 0\cr-\sqrt{6}\end{array}\right)V,
\ee
where $\tilde\sigma$ is the canonically normalized $\sigma$-field. The derivatives along the imaginary directions all vanish. The second derivatives are block diagonal and the block with the real parts of the fields is
\be
\partial_{\phi,s,\tilde\sigma}^2V
= \left(\begin{array}{ccc}
\eta&0&-2\sqrt{3\epsilon}\cr
0&3+\epsilon-{ K}_{S\bar S S\bar S}&  \frac{3}{\delta} \cr
-2\sqrt{3\epsilon}&\frac{3}{\delta} &8+\frac{3}{\delta^2} \end{array}\right)V	 ,
\label{eqn:ddV}
\ee
The mixing between $\phi$ and $\tilde \sigma$ can be neglected when $\epsilon \ll1\ll 1/\delta$ and will be dropped from now on. The perturbation to the inflaton path is then 
\begin{align}
\delta s &= \frac{\sqrt{6}\delta}{{ K}_{S\bar S S\bar S}-\epsilon}+\mathcal O(\delta^2) \ ,\cr
\delta \sigma &=\frac{2\sigma_0\delta^2}{3}\left(1-\frac{3}{{ K}_{S\bar S S\bar S}-\epsilon}\right) +\mathcal O(\delta^3)\  .
\label{eqn:displacement}
\end{align}

One may start the investigation of the inflationary regime for the simplest case, $ K_{S\bar S S\bar S}=0$, which was studied in detail  in \cite{Davis:2008fv} for the case of a quadratic potential, i.e. for  $W = m S\Phi$, see the first plot in Figure \ref{fig:CombinedPotential}. In this case, $\delta s$ is suppressed by $\delta/\epsilon$ (instead of $\delta$) and can thus be $O(1)$ even in the region where $\delta\ll1$. As a consequence, the second order expansion of the potential used to find (\ref{eqn:displacement}) cannot be trusted and the corrections to the effective inflation potential can be big.  This is not necessarily a problem and inflation can be successful \cite{Davis:2008fv}. It does, however, make analytic treatment hard as the KL- and inflation-sectors cannot be disentangled. As a result, for each new set of parameters, one should check numerically whether the field $S$ is light, which may lead to isocurvature/curvaton perturbations of metric. In addition, in the absence of stabilization, the field range where inflation may happen is rather limited, see the first plot in Figure \ref{fig:CombinedPotential}.

Meanwhile, in the case when $\delta\ll 1$ and $K_{S\bar S S\bar S}=\mathcal O(1)$, the deviations $\delta s$ and $\delta \sigma$ are very small and the expressions (\ref{eqn:displacement}) can be trusted. The range of stability can be further increased by making the KL-barrier higher, easiest done by $A\rightarrow C A$ and $B\rightarrow C B$. This makes the barrier higher by a factor of $C^2$. For a quadratic potential this increases the range of stability by a factor $C$. This is demonstrated in the lower plot in Figure \ref{fig:CombinedPotential} for   $W = m S\Phi$. We took  $K_{S\bar S S\bar S}=-4$ and increased the parameters $A$ and $B$ in  (\ref{eqn:paramset}) by a factor of 5. The mass parameter $m$ (which is not the inflaton mass) is taken to be  $m=2.4\times10^{-4}$ so that the full potential $|f(\phi/\sqrt{2})|^2/8\sigma_0^3$, is consistent with the COBE normalization. As we see, the inflationary trajectory is well stabilized. The field $\sigma$ is practically unchanged during the last 60 e-folds of inflation. Inflation may happen for the fields $\phi$ well above $1000$. This means that the slow roll eternal chaotic inflation scenario \cite{Linde:1986fd} can be realized in this model. 
 
Next we turn to the transverse masses. To lowest order in $\delta$ they are
\ba
m_\beta^2&=&\left[2(1-K_{S\bar S \Phi\bar \Phi})+2\epsilon-\eta\right]V \ , \nonumber \\
m_s^2&=&m_\alpha^2=(\epsilon-K_{S\bar SS\bar S})V \ , \nonumber \\
m_\sigma^2&=&m_{\text{Im}\rho}^2=3V/\delta^2=\frac{2}{9}\sigma_0 W^2_{\rho,\rho}(\sigma_0) \ .
\ea
These are identical to those in the decoupled limit discussed in the earlier sections. This is quite surprising since $\partial_s^2V$ differs from (\ref{masss}) by the term $3V$ (which is due to (\ref{eqn:NoScale})). However, looking at the two last rows/columns in (\ref{eqn:ddV}) we see that this term, together with the $1/\delta$ enhanced terms combine into
\be
\left(\begin{array}{cc}3&\frac{3}{\delta}\cr \frac{3}{\delta}&\frac{3}{\delta^2}\end{array}\right)V=\frac{3}{\delta^2}\left(\begin{array}{c}\delta\cr 1\end{array}\right)\left(\begin{array}{cc}\delta&1\end{array}\right)V .
\ee
The new term in $\partial_s^2V$  is thus absorbed into a slight rotation of the heavy $\sigma$ state. The masses should be calculated at the point (\ref{eqn:displacement}) but the correction coming from this is subdominant. We thus conclude that the KL- and inflation-sectors effectively decouple in the limit where $\delta\ll1$ and $ K_{S\bar S S\bar S}=\mathcal O(1)$.  Therefore, for the investigation of the observational consequences of this model, one can simply use the analytical results obtained in \cite{Kallosh:2010ug,Kallosh:2010xz} for the simple supergravity model involving only the fields $S$ and $\Phi$, ignoring the evolution of the volume modulus in the KL model.

\section{Reheating}

Reheating after inflation often can be divided into several qualitatively different stages. Depending on the choice of the inflationary model, reheating may begin with a stage of preheating, a non-perturbative regime of parametric resonance, which rapidly converts the energy of the inflaton field to energy of other particles and classical waves \cite{Kofman:1994rk}. However, eventually this stage ends while some energy still remains stored in the oscillating inflaton field. When the amplitude of this field becomes small enough, one can use the elementary approach to the theory of reheating which describes the perturbative decay of inflaton particles where the reheating temperature after this decay is given by $T_{R} \sim \sqrt{\Gamma}$, and $\Gamma$ is the inflaton decay rate, in Planck units \cite{Dolgov:1982th}.

Perturbative reheating is efficient and can lead to the complete decay of the inflaton field only if this decay continues at a constant rate $\Gamma$ in the limit when the amplitude of the oscillations of the inflaton field vanishes. In other words, it should be a decay process $\phi \to {\it anything}$ rather than some interaction $\phi + \phi \to {\it anything}$. This condition is not automatically satisfied in our class of models. Suppose, for example, that in addition to the fields $\Phi$, $S$ and $T$, we also have matter fields $y$. Consider the diagrams $\phi \to y + y$. The corresponding interaction constant is proportional to $\partial _{\phi,y,y}V$ at the minimum of the potential. This is a straightforward calculation and we find
\be
\partial _{\phi,y,y}V = \frac{1}{2 \sigma_0^3} f(\phi) f'(\phi) ,
\label{phiyy}
\ee
 when evaluated at the minimum of $V(\phi)$ with $S = 0$. But in our model $V(\phi) \sim f(\phi)^{2}$, so at the minimum of the potential in a (nearly) Minkowski vacuum, the decay constant for the process $\phi \to y + y$ vanishes.
 
We next check the coupling of the inflaton to Standard Model fermions.
Starting with the chiral fermion mass matrix,
\be
\- e^{G/2}\bar{\chi^{i}}\left(G_{ij}+G_{i}G_{j}-G_{ij {\bar m}}G^{n \bar m} G_{n}\right){\chi^j} ,
\ee
which becomes
\be
-e^{K/2}\bar{\chi^i} \left(W_{ij}-\frac{2}{3}\frac{W_iW_j}{W}\right) {\chi^j} ,
\ee
when the Goldstino component is subtracted out. 
But because $K_\Phi = W_\Phi = 0$ at the minimum
and we have assumed $W_{i\Phi} = 0$, there are no direct decays of the inflaton to chiral fermions.
A similar argument pertains to the coupling of the inflaton to a scalar-fermion-gaugino. 

In more conventional minimal supergravity models, there is in fact a minimal decay rate for the
inflaton through 3-body gravitational decays \cite{nos,Endo:2006qk} and places constraints on
inflationary models from the overproduction of gravitinos \cite{KTY,grasca}.
Indeed, our computation of the direct decay of the inflaton to matter includes gravitational decays but other channels including the decay to gravitinos are possible, and we check these as well.
For example, in the KL model, one can compute the inflaton couplings to the modulus $\sigma$.
 However, we find this coupling is the same as in Eq. (\ref{phiyy}) multiplied by a factor 
 $6/\sigma_0^2$, and hence also vanishes.
 The coupling of the inflaton to gravitinos is given by
 \begin{eqnarray}
&  - \frac{1}{8} \epsilon^{\mu\nu\rho\sigma}
\left(   G_\Phi \partial_\rho \Phi -  
G_{\bar{\Phi}} \partial_\rho \bar{\Phi}
     \right)
   \bar \psi_\mu \gamma_\nu \psi_\sigma \ , & \nonumber \\
 &
   - \frac{1}{8} e^{G/2} \left( G_\Phi \Phi  +
 G_{\bar{\Phi}} \bar{\Phi}
    \right)
   \bar\psi_\mu \left[\gamma^\mu,\gamma^\nu\right] \psi_\nu &  .
   \label{eq:phi2gravitino}
 \end{eqnarray}
 But as before, because $G_\Phi = K_\Phi + W_\Phi/W$,
 and $K_\Phi = W_\Phi = 0$ at the minimum (recall W =$\Delta W \propto m_{3/2}$), there is no 
 contribution from these terms.
 Indeed, it would appear that there are {\em no} decay channels available for inflaton decay.
 This is very reminiscent of the situation in no-scale supergravity \cite{ekoty}.
 
 There is also the possibility that the inflaton can decay to a single gravitino and inflatino.
 This coupling is given by
 \be
 e^{G/2} G_i \bar\psi_\mu \gamma^\mu \chi^i .
 \ee
 Therefore the decay constant for $\phi \to \chi^i + $ gravitino is
 \be
 e^{G/2} G_{i \Phi} = e^{K/2} \left(K_{i \Phi} W + W_{i \Phi} \right) .
 \ee
Decays of an inflaton to a gravitino plus inflatino (${\tilde \phi}$) are suppressed because the first term
is proportional to the gravitino mass ($e^{K/2} W$) and the second term vanishes 
($W_{\Phi \Phi} = 0$).  However, in principle
 decays to a gravitino + an $S$-ino are possible since $W_{S \Phi} = m$. 
 While this decay does not contribute to reheating, it could be problematic because of the generation of gravitinos.  However, for $m_{3/2} \ll m_\phi$, there is a phase space suppression  of order $(m_{3/2}/m_\phi)^2$  \cite{nop}, for this 
 decay due to degeneracy ($m_\phi - m_{\tilde s} \sim m_{3/2}$).  
 
There is one remaining channel to check,  the decay of the inflaton to 
 the stabilizers, $S$. According to (\ref{eqmass}), the masses of the fields $\phi$ and $s$ are equal to each other at the minimum of the potential, and therefore the decay $\phi \to s+s$ (as well as the decay $s \to \phi+\phi$ which we will discuss later) is kinematically forbidden. 

The suppression of the decay probability of the inflaton field is simultaneously a curse and a blessing.  It is a blessing because with the decay suppressed, the reheating temperature can easily satisfy the bound $T \lesssim 10^{8}$ TeV, which is usually required to avoid the cosmological  gravitino problem for a gravitino at the TeV mass scale \cite{gravitino}. It is a curse because this result implies that unless we do something else, there will be no reheating in this model.

As noted above, the absence of a decay route for the inflaton was generic in no-scale
supergravity \cite{ekoty}.  However, there remains a possibility for inflatons to decay to 
gauge fields and gauginos through a coupling in the gauge kinetic function \cite{ekoty,Endo:2006tf}, $h_{\alpha \beta}$. The supergravity Lagrangian terms of interest include
\begin{eqnarray}
& -\frac{1}{4} ({\rm Re}\, h_{\alpha \beta})F^{\alpha}_{\mu \nu}F^{\beta \mu\nu}
+\frac{i}{4}({\rm Im}\, h_{\alpha \beta})\epsilon^{\mu\nu\rho\sigma}F^{\alpha}_{\mu \nu}F^{\beta \rho\sigma}
&
 \nonumber \\
&
+\left(\frac{1}{4}e^{G/2} {h_{\alpha\beta}^*}_{\bar{n}}G^{k {\bar n}}G_k\lambda^{\alpha}\lambda^{\beta}+h.c.\right) ,
\label{gaugino}
 \end{eqnarray}
Indeed, a non-trivial gauge kinetic function is necessary for the generation of gaugino masses.
For example, for a suitable choice $h_{\alpha \beta} =h(\rho) \delta_{\alpha \beta}$
one could find
\be
m_{1/2} \propto m_{3/2} .
\ee
Therefore, we can in addition include the inflaton dependence in $h$
\be
 h_{\alpha \beta} = (h(\rho) + d_{\phi} \Phi) \delta_{\alpha \beta} .
 \label{hab}
\ee
This would have no effect on the gaugino mass as $G_\Phi = 0$.
However, we would induce a coupling to gauge bosons 
\be
-\frac{1}{4} \langle \frac{\partial h_{\alpha\beta}}{\partial \Phi} \rangle \Phi F^{\alpha}_{\mu \nu}F^{\beta \mu\nu}
= -\frac{ d_{\phi}\delta_{\alpha\beta}}{4\sqrt{2}} \phi F^{\alpha}_{\mu \nu}F^{\beta \mu\nu} .
\ee
Similarly we induce a coupling to gauginos, 
\begin{eqnarray}
 & \frac{1}{4} \langle (e^{G/2} {h_{\alpha\beta}^*}_{\bar{\Phi}}G^{k{\bar \Phi}} G_k )_\Phi \rangle \Phi \lambda^{\alpha} \lambda^{\beta}+h.c. & \nonumber \\
& = \frac{1}{4} \langle e^{G/2} {h_{\alpha\beta}^*}_{\bar{\Phi}}G^{\Phi {\bar \Phi}} G_{\Phi\Phi}  \rangle \Phi  \lambda^{\alpha}\lambda^{\beta}+h.c. & \nonumber \\
 & = - \frac{ d_{\phi} \delta_{\alpha\beta}}{4\sqrt{2}} m_{3/2} \phi  \lambda^{\alpha}\lambda^{\beta}+h.c. & ,
\end{eqnarray}
where we have used $e^{G/2} G_{\Phi\Phi} =  -m_{3/2}$ ($W_{\Phi\Phi} = 0$).
Thus, the decay to gauginos is suppressed by a factor of $(m_{3/2}/m_\phi)^2$
relative to the decay to gauge bosons.

Using a non-trivial gauge kinetic function as in Eq. (\ref{hab}),
we can compute the total decay rate into gauge bosons
as
\be
\Gamma(\phi \rightarrow A_\mu A_\mu) \sim 10^{-2}\times 
 d_{\phi}^2\ \frac{m_{\phi}^3}{M_p^2} \ .
\label{grate}
\ee
Note that this decay rate is suppressed by a factor $d^{2}$ as compared to the standard expectation for the  rate of decay of the inflaton field in supergravity $\Gamma(\phi \rightarrow A_\mu A_\mu) \sim 10^{-2} \frac{m_{\phi}^3}{M_p^2}$. This reduces the reheating temperature, which makes it easier to solve the primordial gravitino problem.

To give a particular example, one may consider the simplest chaotic inflation model with $V = m_{\phi}^{2}\phi^{2}/2$ and $m_{\phi} \sim 6\times 10^{{-6}}$, in Plank mass units.
Our results imply the reheating temperature of  order
\be
T_R \sim   d_{\phi} \times 10^{9}\  {\rm GeV} \ .
\ee
As long as $ d_{\phi} \lesssim 10^{-1}$, excessive reheating and the thermal
production of gravitinos will not occur.

Before concluding this section, we would like to discuss what may happen in those versions of our model where the field $S$ remains very light during inflation and its perturbations are generated, in addition to the perturbations of the inflaton field. In this case, at the end of inflation the field $S$ does not exactly vanish, but instead it takes different values in different parts of the universe. One can show that the decay of the field $S$  to matter fields is also suppressed. In particular, the vertex $\partial _{s,y,y}V $ corresponding to decay $S\to y + y$ in the limit $S\to 0$ is proportional to $f(\phi)$, so it vanishes when evaluated at the minimum of $V(\phi)$ with $S = 0$, $f(\phi)=0$. 

If the field $S$ does not decay, its perturbations will result in undesirable isocurvature perturbations of metric. However, just as the inflaton field, the field $S$ may decay due to the interaction to vector fields with an analogous coupling constant $d_{s}$. If 
$d_{s} \ll d_{\phi}$, the field $S$ may decay much later than the inflaton field, as in the curvaton scenario \cite{curva,Demozzi:2010aj}. In this case the perturbations of the field $S$ generated during inflation produce adiabatic perturbations of metric. Under certain conditions, these perturbations of metric  may be non-gaussian, which may provide additional flexibility to fit new and coming observational data.

However, the curvaton scenario requires many additional nuts and bolts \cite{curva,Demozzi:2010aj,Linde:2005yw}. For example, it could lead to a significant non-gaussianity under the condition that  the amplitude $\delta S/|S|$ is relatively large. To make this condition compatible with the smallness of the perturbations of metric, one may require, e.g., that the main contribution to the density of the universe at the time of the decay of the field $S$ is given not by the classical field $S$, but by the $S$-particles produced during reheating \cite{Demozzi:2010aj,Linde:2005yw}. But in the simplest version of our model the inflaton field does not decay to $S$-particles. One can construct models where such a decay is possible due to non-perturbative effects at the very end of inflation or at the early stages of reheating, e.g. due to a temporary destabilization of the field $S$ at the end of inflation. This can be achieved by a proper choice of the \K\, potential, see \cite{Ferrara:2010in} for a discussion of a closely related regime. Thus, one  can  implement  the curvaton scenario in our model, but it requires additional specification and tuning of its parameters.
 
Meanwhile in the regime  $d_{s} \gtrsim d_{\phi}$, which is perhaps more natural, the field $S$ decays before or shortly after the decay of the inflaton field. In this regime, the contribution of the field $S$ and the products of its decay to the energy density of the universe typically remains much smaller than the corresponding contribution of the inflaton field. In this case, the perturbations of the field $S$ do not lead to significant perturbations of metric, so one can safely ignore the perturbations of the field $S$ even if its mass is much smaller than $H$ during inflation. Therefore in this regime the perturbations of metric are correctly described by the theory of the single inflaton field $\phi$.

 \section{Conclusions}
In this paper, we incorporated the recently developed class of models of chaotic inflation in supergravity \cite{Kallosh:2010ug,Kallosh:2010xz} into a theory of supersymmetry breaking.  We analyzed the possibility of introducing  low-scale SUSY breaking based on the KL model of vacuum stabilization in string theory \cite{Kallosh:2004yh} (see also  \cite{Davis:2008fv}). This mechanism does not require the introduction of additional moduli fields responsible for SUSY breaking, such as Polonyi fields. This solves the cosmological moduli problem which plagues many cosmological models based on supergravity. Furthermore, this mechanism of SUSY breaking has certain distinguishing features which can be tested experimentally.

The models of Refs. \cite{Kallosh:2010ug,Kallosh:2010xz} describe inflation with an arbitrary inflaton potential, which corresponds to a flat direction in the \K\ manifold. In general, unification of these models with the models describing volume modulus stabilization in string theory is rather nontrivial. The large energy density of the inflaton field may affect other scalars (the moduli) and bend or even eliminate the flat direction of the potential. This may make inflation impossible \cite{Kallosh:2004yh}, or at least make it much more complicated, requiring a detailed numerical investigation of the simultaneous evolution of many different scalar fields \cite{Davis:2008fv}.

Fortunately, our results show that in the class of models of Refs.  \cite{Kallosh:2010ug,Kallosh:2010xz} unified with the KL model one can stabilize the inflationary trajectory and preserve the potential along the inflationary direction for a very large range of the values of the inflaton potential $V(\phi)$. In other words, one can reach a certain decoupling, when the KL scenario takes care of the volume modulus stabilization in string theory, as well as the low-scale SUSY breaking, whereas all observational consequences of inflation remain the same as in the simple supergravity models of Refs. \cite{Kallosh:2010ug,Kallosh:2010xz}.

Finally, we analyzed reheating after inflation in this scenario. We have shown (see also \cite{ekoty,Endo:2006tf,Endo:2007sz,Takahashi:2010uw}) that the standard mechanism of reheating with the decay rate $\Gamma \sim m^{3}_{\phi}$ does not work in this class of inflationary models because the flatness of the inflationary potential is related to the vanishing of the \K\, potential along the inflationary trajectory. However, the inflatons can decay to gauge fields and gauginos through a coupling in the gauge kinetic function. This leads to a natural suppression of the decay rate and simplifies the solution of the cosmological gravitino problem.

In conclusion, we constructed a broad class of string theory inspired models of chaotic inflation in supergravity, with a functional freedom of choice of the inflaton potential and the \K\, potential. These models may describe low scale SUSY breaking and they do not suffer from the cosmological moduli and gravitino problems.

The authors are grateful to M. Postma and M. Yamaguchi for enlightening discussions. This work by R.K, A.L. and T.R. was supported by NSF grant PHY-0756174. T.R. is a William R. and Sara Hart Kimball Stanford Graduate Fellow. The work of K.A.O. was supported in part by DOE grant DE-FG02-94ER-40823 at the University of Minnesota. K.A.O. also thanks SLAC (supported by the DOE under contract number DE-AC02-76SF00515) and the Stanford Institute for Theoretical Physics for their hospitality and support while this work was being formulated.

\end{document}